\documentclass{optica-article}

\journal{optica}

\articletype{Research Article}

\usepackage{lineno} 

\usepackage{orcidlink} 

\hypersetup{colorlinks=true,linkcolor=blue,citecolor=blue} 

\usepackage{bm} 

\usepackage[numbers,square,compress]{natbib} 

\usepackage{xcolor} 

\usepackage{pdfpages} 

\usepackage{pgffor} 

\let\Re\relax 
\DeclareMathOperator{\Re}{Re}
\let\Im\relax 
\DeclareMathOperator{\Im}{Im}

\providecommand{\pnl}[1]{{(#1)}} 

\def\supplementfilename{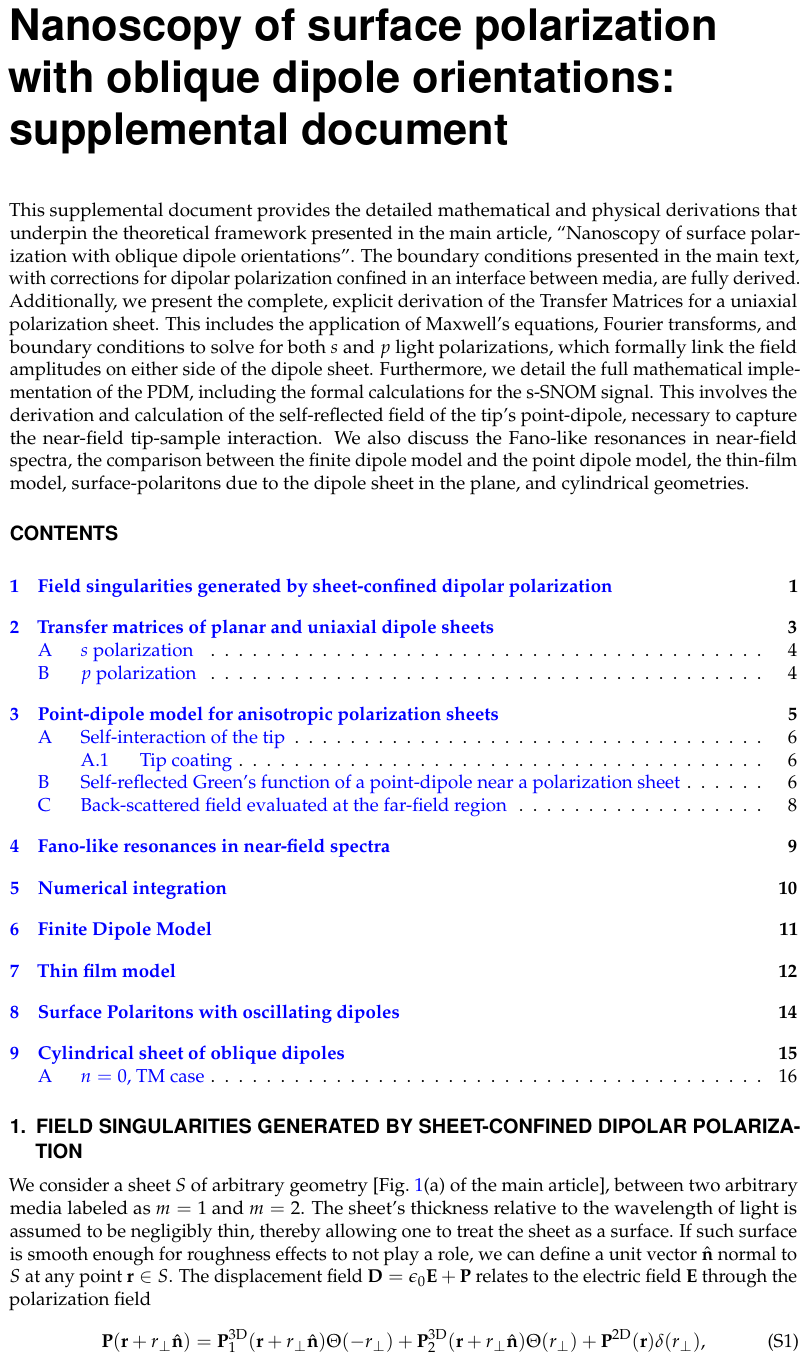}

\pdfximage{\supplementfilename}
\def\numbersupplementpages{\the\pdflastximagepages}

\newif\ifarXiv
\arXivtrue 

\begin{document}

\title{Nanoscopy of surface polarization with oblique dipole orientations}

\author{V. G. M. Duarte,\orcidlink{0009-0009-5836-6084}\authormark{1,2,*} D. A. Miranda,\orcidlink{0000-0002-9784-3340}\authormark{3} D. F. P. Cunha,\orcidlink{0000-0003-4551-9343}\authormark{2,5}  M.~I.~Vasilevskiy,\orcidlink{0000-0003-2930-9434}\authormark{2,5} N.~Asger~Mortensen,\orcidlink{0000-0001-7936-6264}\authormark{3,4} A.~J.~Chaves,\orcidlink{0000-0003-1381-8568}\authormark{1,5} and N. M. R. Peres\orcidlink{0000-0002-7928-8005}\authormark{2,3,5}}

\address{\authormark{1}Department of Physics, Aeronautics Institute of Technology, 12228-900, São José dos Campos, SP, Brazil\\
\authormark{2}International Iberian Nanotechnology Laboratory (INL), Av Mestre Jos\'e Veiga, 4715-330 Braga, Portugal\\
\authormark{3}POLIMA---Center for Polariton-driven Light--Matter Interactions, University of Southern Denmark, Campusvej 55, DK-5230 Odense M, Denmark\\
\authormark{4}Danish Institute for Advanced Study, University of Southern Denmark, Campusvej 55, DK-5230 Odense M, Denmark\\
\authormark{5}Physics Center of Minho and Porto Universities (CF-UM-UP) and Department of Physics, University of Minho, P-4710-057 Braga, Portugal}

\email{\authormark{*}vgmduarte@gmail.com}

\begin{abstract*}

The boundary conditions imposed by confined dipoles with arbitrary orientation on surfaces are presented, extending the conventional in-plane (IP) and out-of-plane (OOP) treatments, here applied for planar and cylindrical sheets. Examples include van der Waals heterostructures, thin films of molecular aggregates, and metal-dielectric interfaces. 
For large dipole strengths, the reflectance peak associated with the dipole oscillation frequency splits into two, revealing the presence of oblique dipoles. The loss function for the dipole sheet reveals pairs of polaritonic resonances originating from the IP and OOP dipole components, accessible through near-field probes. The point-dipole model for s-SNOM shows two distinct peaks, revealing higher sensitivity to dipole obliqueness than reflectance experiments. We apply the model to monolayer WSe$_2$, showing that the oblique dipole formulation with a dipole angle of $5.7^\circ$ yields significant improvements in the qualitative and quantitative agreement with an experiment reported in the literature.
This work proposes a unified language for the description of two-dimensional materials, thin films, and interfaces with anisotropic dipolar responses and shows that near-field methods, sensitive to high in-plane momenta, are suitable for measuring such oblique dipoles.

\end{abstract*}

\section{Introduction}\label{Sec1}

The advances of optical microscopy allowed matter to be probed with light at reduced spatial and temporal scales. In particular, scattering-type scanning near-field optical microscopy (s-SNOM)~\cite{richards2004nearField} was successfully applied to various systems, including semiconductors~\cite{chen2019modern}, organic~\cite{rao2020afm}, and biological samples~\cite{subramaniam1998cell}, revealing intrinsic properties of matter~\cite{atkin2012nanoOptical}. For example, the evanescent component of the electric field originated in the s-SNOM tip can be used to excite surface plasmon-polaritons~\cite{zayats2003nearField}, phonon-polaritons~\cite{huber2008local}, and exciton-polaritons~\cite{hu2019imaging}. Additionally, synchrotron facilities can be used as a source for synchrotron infrared nanospectroscopy~\cite{barcelos2020probing}.

A natural starting point for studying light-matter interactions is the electric dipole~\cite{jackson2021classical}, which represents the first-order response of a neutral charge distribution in the presence of an external field. Frenkel excitons, Wannier--Mott excitons, and quantum surface states, hosted respectively by molecules, semiconductors, and metals, interact with light through their electric dipoles. Frenkel excitons are found in organic materials, and have dipole moments oriented following the molecule orientation~\cite{Bricks2017}. Wannier--Mott excitons play the leading role in two-dimensional (2D) materials, and can emit light whose polarization lies within the 2D material plane or perpendicular to it~\cite{Wang2017}. At last, metallic quantum surface states behave as a series of multipolar corrections to induced charge. These corrections are described through Feibelman's surface response formalism (SRF)~\cite{Feibelman1982}, where the parameter $d_\perp$ represents the out-of-plane component of the induced dipole near the metal surface, originating from the charge distribution of the quantum surface states.

The description of light interaction with excitons is presented in the seminal work of Hopfield~\cite{hopfield1958theory}. However, as pointed elsewhere~\cite{Wang2018}, the dielectric screening reduces both the exciton binding energy and oscillator strengths, making the presence of excitons negligible in most systems~\cite{haug2009quantum}. This can be bypassed in nanostructures such as quantum dots~\cite{Takagahara1993}, or 2D materials~\cite{Wang2018}. The latter have garnered great attention in recent years due to their optical properties and tunability via environment or layer stacking in van der Waals (vdW) heterostructures. For monolayers of MoS$_2$ and other transition metal dichalcogenides (TMDs), the optical properties in the visible range are dominated by the in-plane (bright) exciton peak, which is split into two by virtue of spin-orbit coupling. 2D material monolayers also support out-of-plane (dark) excitations, which are optically a couple orders of magnitude weaker than their in-plane counterparts~\cite{Guilhon2019}. Nonetheless, the optical response of out-of-plane excitons can be probed by polarization-resolved photoluminescence~\cite{Zhou2017,Wang2017} or near-field spectroscopy~\cite{KimKim2021}. Such out-of-plane excitons can also have their emission enhanced by the Purcell effect when coupled with plasmonic cavities~\cite{Bao2023}. Furthermore, vdW heterostructures can support interlayer excitons~\cite{Feng2024}, which can possess static permanent dipoles~\cite{li2020dipolar} and out-of-plane transition dipoles~\cite{Yu_2018}. The static permanent dipoles of interlayer excitons in vdW heterostructures are responsible for an out-of-plane Stark shift~\cite{Jauregui2019} and dipole-dipole interactions~\cite{li2020dipolar}. When the orbital angular momentum component of the band edges is zero in the direction perpendicular to the 2D plane, the polarization of the photons is perpendicular to the plane~\cite{Terry_2018}. This can be used to design vdW heterostructures with out-of-plane photon emission~\cite{Ubrig2020}. Strong out-of-plane photon absorption and emission were observed in monoelemental, binary 2D materials and vdW heterostructures~\cite{Wu2025}.

Besides 2D materials, organic molecules can also exhibit exciton formation. In particular, some molecule species can stack in various geometries and lead to delocalized exciton formation across several molecules~\cite{Bricks2017,Wrthner2011}. Contrary to 2D materials, organic molecular aggregates are highly disordered, but their excitons have dipole moments with fixed and well defined orientation with respect to the aggregate axis. Compared to their monomer counterparts, these molecular aggregates lead to narrower, redshifted resonances and stronger absorption, associated to the formation of large dipole moments that surpass those found in 2D materials~\cite{MISAWA1994251,Kobayashi01051998} and can lead to exciton-photon coupling potentially reaching the ultrastrong coupling regime~\cite{Forn-Diaz2019,Duarte2025}. In fact, J-aggregates were one of the first systems to exhibit superradiance under realistic experimental conditions~\cite{Spano1989}. The description of excitons in such organic materials, as well as their interaction with light, dates back to Kasha's dimer model~\cite{Kasha1,Kasha2}, which describes the fundamental mechanism behind the exciton-exciton coupling and the redshifted resonances. Low-dimensional J-aggregate layers have been employed in films~\cite{Kim2021} and core-shell nanoparticles~\cite{Fofang2008}.

At last, in charged metals the classical assumption is that the charge is confined to the surface, typically forming an interface with vacuum or a dielectric. However, Feibelman and coworkers proposed a multipole series expansion of the induced charge at an interface~\cite{Feibelman1982}. Effectively, the first-order term of this correction corresponds to a dipole-polarized sheet and is responsible for the observation of nonlocal electrodynamic effects in diverse nanophotonics systems~\cite{NonlocalityRoadmap}. Feibelman's SRF introduces the so-called $d$-parameters, related to charge and current density centroid positions defined relative to the metal-dielectric interface. They account for charge spill-out, as well as nonlocal screening, leading to corrections in plasmon dispersion that can be measured. For this reason, Feibelman's SRF underpins the current understanding of mesoscopic electrodynamics in the field of quantum plasmonics, and provides a perspective into the physics of polarized interfaces between media~\cite{Christensen2017,Yang2019}.

In all three systems summarized above, the description of dipoles at surfaces typically considers purely in-plane (IP) or out-of-plane (OOP) orientations. Nonetheless, a theory for surface confined dipoles pointing at oblique directions has been made several decades ago, in the context of monolayers of strongly polarized molecules~\cite{Heinz1983,Felderhof1987Linear,Felderhof1987Electromagnetic}. This theory shows that the continuity of the electric field tangential to a surface is broken when an excess polarization perpendicular to the surface is present, an atypical result that is commonly absent in electrodynamics textbooks~\cite{jackson2021classical}. Despite the generality of this discontinuity, until now it has only flourished in the context of higher order nonlinearities~\cite{Lbau1997,NireekshanReddy2017,Dremetsika2017}. The equation for the jump discontinuity of the tangential electric field is in agreement with Feibelman's SRF, truncated at first-order. The key unifying idea is that oblique dipole polarization confined to a surface, henceforth denominated as a dipole sheet, generates a singularity in the electric field perpendicular to the sheet. In this paper, we build upon the theory of Refs.~\cite{Heinz1983,Felderhof1987Linear,Felderhof1987Electromagnetic}  with a description of the dipole sheet in terms of a surface of zero thickness, here applied to planar and cylindrical sheets, that can be straightforwardly extended to smooth curved geometries.

Although several works discuss the properties of IP and OOP dipole sheets, such as excitons, the presence of oblique dipoles generates intrinsic IP-OOP coupling and interference effects, which have not been analyzed before. One possible system in which this effect can be important and measurable is TMDs alloys. TMDs typically have OOP dark and IP bright exciton states close in energy. For a MoX$_2$ monolayer (X=S,Se), the lowest excitonic state is bright, while for WX$_2$, the lowest excitonic state is dark. 
Therefore, there is a crossing between the bright and dark exciton energy states for some $x$ in the alloy Mo$_x$W$_{1-x}$X$_2$~\cite{AlloyTMDs}, effectively resulting in oblique dipoles. 
A strong out-of-plane electric field can excite dark exciton states~\cite{Park2018}, and it is also known that bright excitons can be probed by near-field techniques~\cite{Zhang2022}. Thus, techniques such as s-SNOM will simultaneously excite and probe the dark and bright excitons. In this work we focus on discussing what is expected from such measurements.

The paper is organized as follows: in Sec.~\ref{sec:boundary_conditions} we present the general electromagnetic description of dipole polarization confined to a surface of deeply subwavelength thickness. In Sec.~\ref{sec:reflection} we apply this theory to the reflection of a uniaxial excitonic sheet, and discuss the physical features arising from a coexistence of finite IP and OOP dipole components. Sec.~\ref{Sec4} improves further by providing a model for near-field spectroscopy of uniaxial excitonic sheets. The near-field spectral features coming from dipole obliqueness are discussed, devising a possible technique for characterization of oblique and OOP dipoles in 2D materials. The main conclusions are reiterated in Sec.~\ref{Sec5}.

\section{Sheet-confined dipolar polarization}\label{sec:boundary_conditions}

Discontinuities on the electromagnetic fields caused by accumulation of dipolar polarization in a thin sheet have been the subject of a number of both modern and legacy studies~\cite{Feibelman1982,Heinz1983,Felderhof1987Linear,Felderhof1987Electromagnetic,Yang2019,Hansen2026}. The OOP degree of freedom is responsible for strong non-local deviations in the optical response of photonic systems with thickness deeply below the wavelength of light. However, the role of OOP polarization remains poorly understood, as studies of IP-isotropic 2D systems usually neglect the presence of finite OOP optical response \cite{Hansen2026}. Here, we devise an electromagnetic model for dipolar polarization confined at a thin sheet, treated as a surface of zero thickness. In this picture, the polarization accumulated in the sheet generates singularities in the electromagnetic sources, responsible for the observed field discontinuities.

\begin{figure}[ht]
    \centering
    \includegraphics{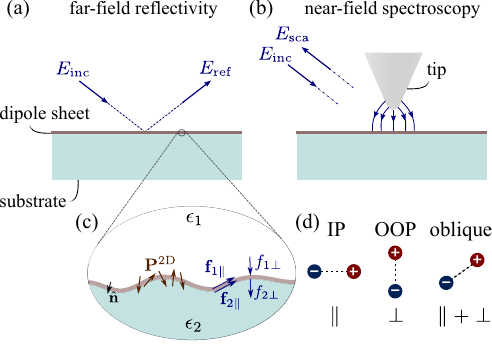}
    \caption{{Schematic representation of a dipole sheet in the optical configurations considered in this work. \pnl{a}~Plane wave fields impinge on the sheet, on top of a transparent substrate. The sheet reflectivity is then analyzed by taking the ratio of the reflected field ($E_\mathrm{ref}$) to the incident field ($E_\mathrm{inc}$). \pnl{b}~A metallic tip in close proximity to sheet oscillates generating harmonic patterns in the electromagnetic field that propagate to the scattered field ($E_\mathrm{sca}$). The tip-sheet coupling is retrieved through the high-order harmonics of $E_\mathrm{sca}$, upon proper normalization. \pnl{c}~Zoomed view of the dipole sheet, showing the surface polarization field $\mathbf{P}^\mathrm{2D}$ of the dipole sheet and the fields on each sheet side (with $\mathbf{f}=\mathbf{D},\mathbf{B},\mathbf{E},\mathbf{H}$). The media above and below the sheet have permittivities $\epsilon_{1,2}$, and $\hat{\mathbf{n}}$ represents a unit vector normal to any point in the sheet, treated as a surface of zero thickness. (d) IP, OOP and oblique dipole markers used in the remaining figures of this work.}}
    \label{fig:optical_configurations}
\end{figure}

The boundary conditions (BCs) read as
\begin{subequations}\label{eq:BCs}
\begin{align}
    D_{2,\perp}-D_{1,\perp}&=-\nabla_\parallel\cdot\mathbf{P}^\mathrm{2D}_\parallel,
    \\
    B_{2,\perp}-B_{1,\perp}&=0,
    \\
    \mathbf{E}_{2,\parallel}-\mathbf{E}_{1,\parallel}&=-\dfrac{\nabla_\parallel P^\mathrm{2D}_\perp}{\epsilon_0},\label{eq:BC Eparallel}
    \\
    \mathbf{H}_{2,\parallel}-\mathbf{H}_{1,\parallel}&=-i\omega\mathbf{P}^\mathrm{2D}\times\hat{\mathbf{n}},
\end{align}
\end{subequations}
where we have used the normal unit vector $\hat{\mathbf{n}}$, as indicated in Fig.~\ref{fig:optical_configurations}\pnl{c}, and the normal coordinate $r_\perp$. The fields are evaluated at $\mathbf{r}+r_\perp\hat{\mathbf{n}}$, for any point $\mathbf{r}$ on the zero-thickness surface representing the dipole sheet. The index $m=1$ ($m=2$) is shorthand for field evaluations at $r_\perp=0^-$ ($0^+$); the index $\parallel$ ($\perp$), for the IP (OOP) field components. The oblique dipole then follows as a combination of
IP and OOP dipole components [Fig.~\ref{fig:optical_configurations}\pnl{d}]. Upon proper definition of a surface polarization field $\mathbf{P}^\mathrm{2D}$, these BCs have been used to model finite OOP optical response in strongly polarized molecular layers~\cite{Heinz1983,Felderhof1987Linear}, 2D materials~\cite{Hansen2026}, and even metal-dielectric interfaces~\cite{NireekshanReddy2017,Yang2019}. One can re-derive the BCs~\eqref{eq:BCs} by postulating the validity of Maxwell's equations for singular fields and applying closed-surface and -contour integrations. More importantly, we show that the dependence of the BCs on the surface-polarization field $\mathbf{P}^\mathrm{2D}$ is universally a consequence of the field singularities (see Sec.~1 of the Supplement 1)
\begin{subequations}
\begin{align}
    \int_{0^-}^{0^+} dr_\perp \hat{\mathbf{n}}\cdot\mathbf{E}(\mathbf{r}+r_\perp\hat{\mathbf{n}})&=-\frac{1}{\epsilon_0}\hat{\mathbf{n}}\cdot\mathbf{P}^\mathrm{2D},
    \\
    \int_{0^-}^{0^+} dr_\perp \hat{\mathbf{n}}\times\mathbf{D}(\mathbf{r}+r_\perp\hat{\mathbf{n}})&=\hat{\mathbf{n}}\times\mathbf{P}^\mathrm{2D},   
\end{align}
\end{subequations}
which leaves finite contributions on the field integrals over the lateral surface of an infinitesimal Gaussian pillbox (as well as on the lateral line segment of an infinitesimal closed contour) around the sheet. The displacement field reads as $\mathbf{D}=\epsilon_0\mathbf{E}+\mathbf{P}$, with
\begin{equation}\label{eq:P}
    \mathbf{P}(\mathbf{r}+r_\perp\hat{\mathbf{n}})=\mathbf{P}^{(0)}(\mathbf{r}+r_\perp\hat{\mathbf{n}}) +\mathbf{P}^\mathrm{2D}(\mathbf{r})\delta(r_\perp),
\end{equation}
where $\mathbf{P}^{(0)}$ represents the regular (non-singular) contributions to the polarization field. 

The usual constitutive relation of the polarization field ($\mathbf{P}=\epsilon_0\overset{\leftrightarrow}{\chi}\mathbf{E}$ at linear order) becomes ill-defined for a singular polarization field as in Eq.~\eqref{eq:P}. For this reason, a universal form for the constitutive relation between $\mathbf{P}^\mathrm{2D}$ and the electromagnetic fields, especially for the OOP component $P^\mathrm{2D}_\perp$, remains open to debate~\cite{Felderhof1987Linear,Yang2019,NireekshanReddy2017,Hansen2026}. The most general proposal is to write $\mathbf{P}^\mathrm{2D}$ as a weighted-sum of the electric fields on each sheet side, with susceptibility tensors $\overset{\leftrightarrow}{\chi}_{1,2}$ to be determined~\cite{Felderhof1987Linear}:
\begin{equation}\label{eq:P2D}
    \mathbf{P}^\mathrm{2D}=\epsilon_0\overset{\leftrightarrow}{\chi}_1\mathbf{E}_1+\epsilon_0\overset{\leftrightarrow}{\chi}_2\mathbf{E}_2.
\end{equation}
For a sheet between uniaxial dielectrics, this translates into (denoting the identity tensor as $\overset{\leftrightarrow}{1}$) $\overset{\leftrightarrow}{\chi}_1=\epsilon_{1,\parallel}d_\parallel(\overset{\leftrightarrow}{1}-\hat{\mathbf{n}}\otimes\hat{\mathbf{n}}) - d_\perp\hat{\mathbf{n}}\otimes\hat{\mathbf{n}}$ and $\overset{\leftrightarrow}{\chi}_2=-\epsilon_{2,\parallel}d_\parallel(\overset{\leftrightarrow}{1}-\hat{\mathbf{n}}\otimes\hat{\mathbf{n}}) + d_\perp \hat{\mathbf{n}}\otimes\hat{\mathbf{n}}$, where the Feibelman's $d$--parameters encompass the microscopic nature of charge fluctuations in the sheet~\cite{Yang2019}.

\begin{table}[htb]
    \centering
    \caption{\label{tab:bc}Boundary conditions (BCs) {with corrections for surface-normal polarization, showing the emergence of a discontinuity in the tangential electric field $\mathbf E_\parallel$. To keep the equations compact, the notation ${[f]}=f_2-f_1$ was used, for a vector (or scalar) field $f$.}
    {To illustrate two physical realizations of this BCs, we show how they apply to a planar layer of strongly polarizable molecules adsorbed in a dielectric~\cite{Felderhof1987Linear}, and a metal-dielectric interface with charge fluctuations described by Feibelman's $d$--parameters~\cite{Yang2019}. Whenever necessary, the equations reproduced from Refs.~\cite{Yang2019,Felderhof1987Linear} were adapted to comply with SI units and our choices of notation.}
    }
    \begin{tabular}{
    >{\centering\arraybackslash}p{1.2in}
    >{\centering\arraybackslash}p{1.3in}
    >{\centering\arraybackslash}p{1.2in}}
    \noalign{\hrule height 1pt}
    {Metal-dielectric interface}~\cite{Yang2019} & {Planar layer of molecules}~\cite{Felderhof1987Linear} & {Generalized BCs (This work)}\\
    \hline\noalign{\vskip 3pt}
    ${[D_{\perp}]} = d_\parallel \nabla_\parallel \cdot {[\mathbf{D}_\parallel]}$ & ${[E_\perp]} = -\dfrac{\nabla_\parallel \cdot {\mathbf{P}^\mathrm{2D}_\parallel}}{\epsilon\epsilon_0}$ & ${[D_\perp]} = -\nabla_\parallel\cdot{\mathbf{P}^\mathrm{2D}_\parallel}$\\
    ${[B_\perp]} = 0$ & ${[B_\perp]} = 0$ & ${[B_\perp]} = 0$\\
    ${[\mathbf{E}_\parallel]} = -d_\perp\nabla_\parallel {[E_\perp]}$ & ${[\mathbf E_\parallel]} = -\dfrac{\nabla_\parallel {P^\mathrm{2D}_\perp}}{\epsilon\epsilon_0}$ & ${[\mathbf E_\parallel]} = -\dfrac{\nabla_\parallel {P^\mathrm{2D}_\perp}}{\epsilon_0}$\\
    ${[\mathbf H_\parallel]} = i\omega d_\parallel {[\mathbf D_\parallel]} \times\hat{\mathbf n}$ & ${[\mathbf{B}_\parallel]} = -i\omega \mu\mu_0{\mathbf{P}^\mathrm{2D}_\parallel} \times \hat{\mathbf z}$ & ${[\mathbf H_\parallel]} = -i\omega{\mathbf{P}^\mathrm{2D}_\parallel} \times \hat{\mathbf n}$\\
    \noalign{\hrule height 1pt}
    \end{tabular}
\end{table}

An important source of ambiguity on the proper interpretation of $\mathbf{P}^\mathrm{2D}$ must be addressed. Other works have suggested that the tangential electric field BC~\eqref{eq:BC Eparallel} must carry an additional factor of $1/\epsilon_\perp$~~\cite{Felderhof1987Linear,NireekshanReddy2017,Hansen2026}. In this picture, the surface polarization field represents solely the intrinsic sources within the sheet, and a factor of $1/\epsilon_\perp$, defined as some combination between the permittivities on each sheet side, needs to be included to account for screening effects. An attempt at splitting different sources of dipole-polarization like this easily runs into complications once the nature of the electromagnetic sources in the sheet are put to doubt. For instance, in 2D material research it is frequently necessary to use hBN encapsulation~\cite{Wang2015,Han2019,Pace2021,Ma2022,Ye2023hBN}, which hybridizes with the target material. Nonetheless, the BCs~\eqref{eq:BCs} remain valid if $\mathbf{P}^\mathrm{2D}$ is interpreted as the \textit{net} field after all these effects are included.

To conclude this Section, in Tab.~\ref{tab:bc} we compare the BCs here presented to those of Felderhof's polarization sheet formalism~\cite{Felderhof1987Linear} and Feibelman's SRF~\cite{Yang2019}.

\section{Far-field reflectance and surface polaritons of uniaxial dipole sheets}\label{sec:reflection}

\subsection{Transfer matrices}\label{sec:reflection.transfer_matrices}

The transfer matrices allow for extraction of optical properties such as the reflection and transmission coefficients, as well as for an extensible way of chaining together layered materials with dipole sheets involved. In this section, we derive the transfer matrices for a dipole sheet under incidence of linearly polarized light, as illustrated in Fig.~\ref{fig:optical_configurations}.

\begin{table}[htb]
\centering
\caption{\label{tab:chi}Application of the uniaxial susceptibility tensors~\eqref{eq:chi uniaxial}, governing the general linear constitutive relation~\eqref{eq:P2D}, to a conventional IP-isotropic 2D material with surface conductivity $\sigma(\omega)$ and a metal-dielectric interface with permittivities $\epsilon_{1,2}$ and Feibelman's parameters $d_\parallel, d_\perp$.}
\begin{tabular}{c|cc}
\noalign{\hrule height 1pt}
& IP-isotropic 2D materials~ & metal-dielectric interface~\cite{Yang2019}
\\
\hline
$\chi_{1,\parallel}$ & $\frac{i}{2\epsilon_0\omega}\sigma(\omega)$ & $\epsilon_1 d_\parallel(\omega)$
\\
$\chi_{2,\parallel}$ & $\frac{i}{2\epsilon_0\omega}\sigma(\omega)$ & $-\epsilon_2 d_\parallel(\omega)$
\\
$\chi_{1,\perp}$ & $0$ & $-d_\perp(\omega)$
\\
$\chi_{2,\perp}$ & $0$ & $d_\perp(\omega)$
\\
\noalign{\hrule height 1pt}
\end{tabular}
\end{table}

We assume a linear constitutive relation for $\mathbf{P}^\mathrm{2D}$ as in Eq.~\eqref{eq:P2D}, with susceptibility tensors
\begin{equation}\label{eq:chi uniaxial}
    \overset{\leftrightarrow}{\chi}_m(\omega) =
    \begin{pmatrix}
        \chi_{m,\parallel}(\omega) & 0 & 0\\
        0 & {\chi_{m,\parallel}}(\omega) & 0\\
        0 & 0 & {\chi_{m,\perp}}(\omega)
    \end{pmatrix}.
\end{equation}
As Eq.~\eqref{eq:chi uniaxial} shows, the OOP anisotropy introduced by $\chi_{m,\perp}\neq\chi_{m,\parallel}$ produces the uniaxial response necessary for characterization of OOP and oblique dipoles. In what follows, to retain as much generality as possible the susceptibility tensors~\eqref{eq:chi uniaxial} will be left generic. Some particularizations of these susceptibility tensors to common optical setups are given in Tab.~\ref{tab:chi}.

{For a dipole sheet flanked by homogeneous isotropic dielectrics $\epsilon_m$, plane waves propagate with wavevectors $\mathbf{k}=\mathbf{k}_\parallel+k_{m,\perp}\hat{\mathbf{n}}$, with $\mathbf{k}_\parallel$ the wavenumber parallel to the interface and $k_{m,\perp}=\sqrt{\epsilon_m(\omega^2/c^2)-k_\parallel^2}$}. The normal vector $\hat{\mathbf n}$ becomes uniform across the entire sheet in the planar case. We define field amplitudes {$F_{m,\lambda,\pm}$,} such that
\begin{subequations}
\begin{equation}
    \mathbf{E}_{m,s}(\mathbf{r})=(F_{m,s,+}e^{i\mathbf{k}_m^+\cdot\mathbf{r}}+F_{m,s,-}e^{i\mathbf{k}_m^-\cdot\mathbf{r}})\mathbf{e}_s,
\end{equation}
for $s$-polarized plane waves, and
\begin{eqnarray}
    \mathbf{B}_{m,p}(\mathbf{r})=-(F_{m,p,+}e^{i\mathbf{k}_m^+\cdot\mathbf{r}}+F_{m,p,-}e^{i\mathbf{k}_m^-\cdot\mathbf{r}})\mathbf{e}_p,
\end{eqnarray}
\end{subequations}
for $p$-polarized plane waves. The vector {$\mathbf e_s$} is defined as ${\mathbf e_s}=\mathbf k_\parallel\times\hat{\mathbf n}/k_\parallel$, with $\mathbf k^\pm_m = \mathbf k_\parallel \pm k_{m,\perp}\hat{\mathbf n}$ {for propagation on the $\pm\hat{\mathbf{n}}$ directions.}

The transfer matrices, defined according to
\begin{equation}
    \begin{pmatrix}
        {F_{1,\lambda,+}}\\
        {F_{1,\lambda,-}}
    \end{pmatrix}
    =
    {T_\lambda}
    \begin{pmatrix}
        {F_{2,\lambda,+}}\\
        {F_{2,\lambda,-}}
    \end{pmatrix},
\end{equation}
are given by the expressions
\begin{subequations}\label{eq:tms}
\begin{align}
    {T_s} &= \frac{1}{2}
    \begin{pmatrix}
        1-{\beta_s}+{\eta_s} & 1-{\beta_s}-{\eta_s}\\
        1+{\beta_s}-{\eta_s} & 1+{\beta_s}+{\eta_s}
    \end{pmatrix},\label{eq:STE}\\
    {T_p} &= \frac{1}{2(1-{\beta_{1,p,\parallel} \beta_{1,p,\perp}})}
    \begin{pmatrix}
        1 - {\beta_{1,p,\parallel}} & 1 - {\beta_{1,p,\perp}} \\
        -1-{\beta_{1,p,\parallel}} & 1 + {\beta_{1,p,\perp}}
    \end{pmatrix}
    \begin{pmatrix}
        {\eta_p} (1 - {\beta_{2,p,\perp}}) & -{\eta_p} (1+{\beta_{2,p,\perp}})\\
        1 - {\beta_{2,p,\parallel}} & 1 + {\beta_{2,p,\parallel}}
    \end{pmatrix}\label{eq:STM},
\end{align}
\end{subequations}
where we defined the auxiliary quantities
\begin{subequations}
    \begin{eqnarray}
        {\eta_s} = \frac{{k_{2,\perp}}}{{k_{1,\perp}}},
        \;\;\;
        {\beta_s} = i\frac{\omega^2}{c^2} \frac{{\chi_{1,\parallel}} + {\chi_{2,\parallel}}}{k_{1,\perp}},
        \\
        {\eta_p} = \dfrac{{\epsilon_1} {k_{2,\perp}}}{{\epsilon_2} {k_{1,\perp}}},\label{eq:auxiliary coefficients of the transfer matrix for s polarization}
        \;\;\;
        {\beta_{m,p,\parallel}} = i \frac{{k_{m,\perp} \chi_{m,\parallel}}}{{\epsilon_m}},
        \;\;\;
        {\beta_{m,p,\perp}} = i\frac{{k_\parallel}^2 {\chi_{m,\perp}}}{{k_{m,\perp}}}.\label{eq:auxiliary coefficients of the transfer matrix for p-polarization}
    \end{eqnarray}
\end{subequations} 
For a detailed derivation of the transfer matrices~\eqref{eq:tms}, the reader is referred to {Section 2 of the Supplement 1}.

The transfer matrices~\eqref{eq:tms} were derived treating the dipole sheet as a zero-thickness surface. An alternative approach is to define a finite effective thickness $d$ and apply the usual transfer matrices of bulk media. In this picture, the sheet's optical response relies on a dielectric function that varies depending on $d$ and the permittivities of the surrounding environments~\cite{Majerus2018electrodynamics}. Nonetheless, both these treatments give the same optical response in the regime where $d$ is deeply below the wavelength of incoming light (see Sec.~7 of the Supplement 1), and for this reason we stick with the zero-thickness surface treatment for its improved transparency on the description of atomically-thin dipole sheets.

\subsection{Fresnel coefficients}\label{sec:reflection.fresnel}

The Fresnel coefficients of reflection ($r_\lambda = F_{1,\lambda,-}/F_{1,\lambda,+}$) and transmission ($t_\lambda = F_{2,\lambda,+}/F_{1,\lambda,+}$) are obtained by setting $F_{2,\lambda,-}=0$ on the transfer matrices~\eqref{eq:STE} and~\eqref{eq:STM}:
\begin{subequations}
\begin{eqnarray}
    {t_s}({k_\parallel},\omega)&=&\frac{2}{1-{\beta_s}+{\eta_s}},
    \\
    {r_s}({k_\parallel},\omega)&=&\frac{1+{\beta_s}-{\eta_s}}{1-{\beta_s}+{\eta_s}},
    \\
    {t_p}({k_\parallel},\omega)&=&\frac{2(1-{\beta_{1,p,\parallel}\beta_{1,p,\perp}})}{(1-{\beta_{1,p,\perp}})(1-{\beta_{2,p,\parallel}})+{\eta_p}(1-{\beta_{1,p,\parallel}})(1-{\beta_{2,p,\perp}})},
    \\
    {r_p}({k_\parallel},\omega)&=&\frac{(1+{\beta_{1,p,\perp}})(1-{\beta_{2,p,\parallel}})-{\eta_p}(1+{\beta_{1,p,\parallel}})(1-{\beta_{2,p,\perp}})}{(1-{\beta_{1,p,\perp}})(1-{\beta_{2,p,\parallel}})+{\eta_p}(1-{\beta_{1,p,\parallel}})(1-{\beta_{2,p,\perp}})}.\label{eq:Fresnel reflection coefficient for p-polarized light}
\end{eqnarray}
\end{subequations}
The reflectance and transmittance follow from {$|r_\lambda|^2$ and $\eta^{(\lambda)} |t_\lambda|^2$, respectively}.

\subsection{Reflectance}\label{sec:reflection.reflectance}

\begin{figure}[ht]
    \centering
    \includegraphics{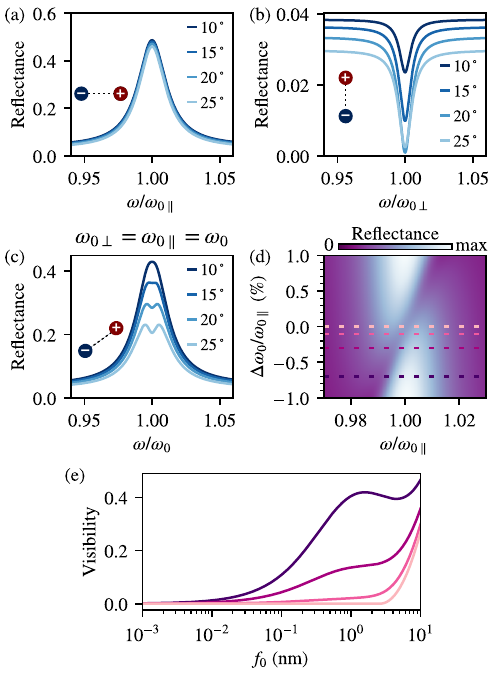}
    \caption{ Reflectance of excitons in a uniaxial sheet. \pnl{a--c}~Reflectance spectrum at different incidence angles for a sheet with \pnl{a}~IP dipoles at frequency $\omega_{0\parallel}$ with oscillator strengths $\{f_\parallel,f_\perp\}=\{\frac{1}{2}f_0\cos^2(30^\circ),0\}$, \pnl{b}~OOP dipoles at $\omega_{0\perp}$ with $\{f_\parallel,f_\perp\}=\{0,f_0\sin^2(30^\circ)\}$, and \pnl{c}~oblique dipoles at $\omega_0=\omega_{0\parallel}=\omega_{0\perp}$ with $\{f_\parallel,f_\perp\}=\{\frac{1}{2}f_0\cos^2(30^\circ),f_0\sin^2(30^\circ)\}$. Parameters: $\omega_{0\parallel}=2.1$\,eV, $f_0=5$\,nm, $\epsilon_1=1$, $\epsilon_2=2.25$, and $\gamma=0.02$\,eV. \pnl{d}~Same as panel \pnl{c}, but with a finite detuning $\Delta\omega_0=\omega_{0\perp}-\omega_{0\parallel}$ and an incidence angle fixed at $\theta_\mathrm{inc}=25^\circ$, where the double peak is most visible. \pnl{e} Visibility, defined as $(P-D)/(P+D)$ in terms of the reflectance peaks $P$ and dips $D$, at fixed $\omega_{0\perp}$ cuts selected from panel~\pnl{d}, as a function of the total oscillator strength $f_0$.
    }
    \label{fig:reflection}
\end{figure}

Figure~\ref{fig:reflection} shows the evolution of the reflectance spectrum as a function of incidence angle for susceptibility functions modeled as the Lorentzians
\begin{subequations}\label{eq:lorentzian}
\begin{align}
    \chi_{m,\parallel}(\omega) &= \dfrac{f_{m,\parallel}\omega_{0\parallel}^2}{\omega_{0\parallel}^2-\omega^2-i\omega\gamma},
    \\
    \chi_{m,\perp}(\omega) &= \dfrac{f_{m,\perp}\omega_{0\perp}^2}{\omega_{0\perp}^2-\omega^2-i\omega\gamma},
\end{align}
\end{subequations}
yielding the usual response of IP (OOP) excitons at resonance frequency $\omega_{0\parallel}$ ($\omega_{0\perp}$). The oscillator strengths $f_{m,\parallel}$ and $f_{m,\perp}$ modulate the sheet's response to incident light in the IP and OOP directions, respectively, and the damping parameter $\gamma$ encompasses all loss sources, thereby modulating the Lorentzian peak's height and width. Throughout most of this work, $\gamma\approx 0.01\omega_{0\parallel}$. For comparison, monolayers of TMDs usually have linewidths in the order of $10$--$20$\,meV, \textit{i.e.}, at most $\sim 1.3\%$ of their optical transition frequencies~\cite{Selig2016ExcitonicLinewidth}. {The values of the oscillator strengths $f_{m,\parallel},f_{m,\perp}$ can vary from $0.2$ nm in TMDs \cite{Zhang2022}  to $29$ nm in J-aggregates \cite{Jaggregate_huge}  (see Sec. \ref{Sec4.2} for the equations that relate the dielectric function to 2D susceptibility description)}. 

In Fig.~\ref{fig:reflection}, the sheet is placed on top of a silica substrate, with permittivity $\epsilon_2=2.25$ and negligible absorption at optical frequencies~\cite{aizpurua2008substrate}. Additionally, the sheet is assumed to respond symmetrically to the fields on both sides, for simplicity [$f_{1,\parallel}=f_{2,\parallel}=f_\parallel$ and $f_{1,\perp}=f_{2,\perp}=f_\perp$ in Eqs.~\eqref{eq:lorentzian}, such that $\chi_{1\parallel}=\chi_{2\parallel}=\chi_\parallel$ and $\chi_{1\perp}=\chi_{2\perp}=\chi_\perp$]. Since we are interested in capturing the combined effect of IP and OOP polarization, $p$-polarization was considered, and we write the oscillator strengths using the polar representation
\begin{subequations}\label{eq:symmetric_oscillator_strengths}
\begin{align}
    f_{\parallel} &= \frac{1}{2} f_0 \cos^2(\theta_\mathrm{dip}),
    \\
    f_{\perp} &= f_0 \sin^2(\theta_\mathrm{dip}),
\end{align}
\end{subequations}
in terms of the total oscillator strength $f_0=2f_\parallel+f_\perp$ and an angle $\theta_\mathrm{dip}$, henceforth called the dipole angle. The reflectance spectrum forms a peak in the IP case~[Fig.~\ref{fig:reflection}\pnl{a}] and a dip in the OOP case~[Fig.~\ref{fig:reflection}\pnl{b}], associated to orientation alignment between the dipoles and the incident field. Near resonance with the exciton, close-to-normal incidence angles favor coupling with the IP dipole component, thus the sheet with IP dipoles strongly radiates back into the reflectance channel, while the sheet with OOP dipoles absorbs most of the incoming light. As the incidence angle increases, however, the incoming light partially decouples from the IP component and increasingly couples to the OOP component. The behavior of the oblique dipole sheet follows from a combination of these two effects: the IP peak is notched down slightly by the OOP dip, forming a pair of lower peaks with a dip in the middle~[Fig.~\ref{fig:reflection}\pnl{c}]. The visibility of this notched peak is highly sensitive to incidence angle and to oscillator strength, which control field-dipole alignment and the strength of the peak notching respectively~[Fig.~\ref{fig:reflection}\pnl{e} for $\omega_{0\perp}=\omega_{0\parallel}$].

At finite OOP-IP exciton detuning, the notched peak evolves to an asymmetric peak-dip pair~[Fig.~\ref{fig:reflection}\pnl{d}]. At sufficiently small detunings, one is expected to encounter a reflectance spectrum nearly identical to that of zero detuning, indicated by the persistence of the double peak feature in Fig.~\ref{fig:reflection}\pnl{d}. Large detunings, conversely, give rise to a decoupling of the peak-dip pair forming a pair of optical signatures which, for a transparent substrate like the one used in Fig.~\ref{fig:reflection}, are almost completely independent from one another. The large increase in visibility observed in Fig.~\ref{fig:reflection}\pnl{e} for finite detunings is then a consequence of measuring a peak against a dip far away on its tail. Nonetheless, a mildly improved visibility is expected at detunings roughly around $0$--$0.5\%$ of the in-plane dipole frequency $\omega_{0\parallel}$, without losing significant IP-OOP coherence.

\subsection{Surface polariton modes}\label{sec:reflection.loss_function}

We follow up on the investigation of the excitonic sheet with oblique dipoles, characterized by the susceptibility functions~\eqref{eq:lorentzian} with symmetric oscillator strengths~\eqref{eq:symmetric_oscillator_strengths}. At the center of our attention is the sheet with detuning of zero or lower than $0.5\%$ of the in-plane dipole frequency, representing either an exciton with a pure oblique dipole or a pair of IP and OOP nearly-degenerate excitons. In this situation, we denote by $\omega_{0\perp}\approx\omega_{0\parallel}=\omega_{0}$ the oscillation frequency of the dipoles.

\begin{figure}[ht]
    \centering
    \includegraphics{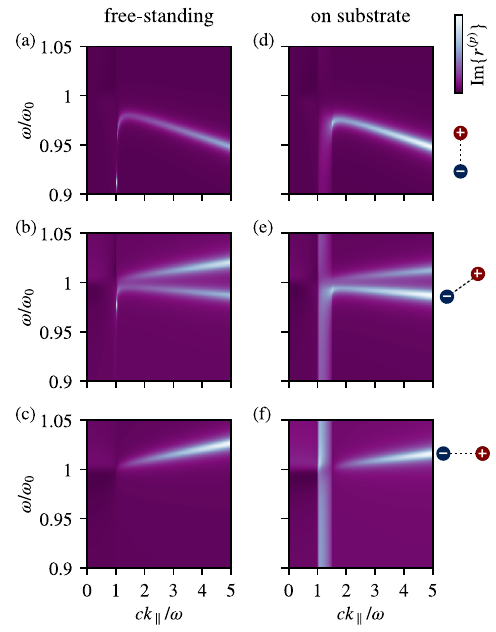}
    \caption{ Loss function of excitons in a uniaxial sheet, with IP and OOP transitions at $\omega_{0\perp}\approx\omega_{0\parallel}=\omega_0$. \pnl{a--c}~Dispersion relation (peaks in the loss function) for a free-standing sheet with \pnl{a}~OOP dipoles ($\theta_\mathrm{dip}=\pi/2$), \pnl{b}~oblique dipoles ($\theta_\mathrm{dip}=\pi/6$), and \pnl{c}~IP dipoles ($\theta_\mathrm{dip}=0$). Parameters: $\omega_0=2.1$\,eV, $f_0=2$\,nm, $\epsilon_1=1$, $\epsilon_2=1$, and $\gamma=0.02$\,eV. \pnl{d--f}~Same as in panels \pnl{a--c}, but for a sheet on top of glass ($\epsilon_2=2.25$).
    }
    \label{fig:loss_function}
\end{figure}

The loss function, defined as $\Im\{r\}$, can be used to infer about surface modes excited by optical means, such as electron energy loss spectroscopy (EELS)~\cite{Goncalves2016}. At the right of the light line $k_\parallel=\omega/c$, the excitonic sheet with oblique dipoles presents two {surface} modes~[Fig.~\ref{fig:loss_function}], above and below the excitonic resonance. The horizontal axes in Fig.~\ref{fig:loss_function} were normalized by $\omega/c$ in order to highlight the degree of confinement of the surface modes. 
{The upper dispersion curve reproduces the behaviour already described in the literature of a polariton mode in a surface with dipoles oscillating in-plane, as is the case of excitons in a single-layer of hexagon boron nitride shown in}~\cite{ferreira2019excitonsHexagonalBoron}. 
{For in-plane excitons, only s-polarized modes have a lower polariton branch with negative group velocity~\cite{ferreira2019excitonsHexagonalBoron}.}
Here, by contrast, $p$-polarized light also excites a lower branch. The insertion of the substrate slightly tilts the polariton branches further away from the exciton line $\omega=\omega_0$, but its qualitative aspect is preserved. The group velocity of the upper branch is strictly positive and varies weakly at large $k_\parallel$, which is a feature of the chosen susceptibility model [see Eq.~\eqref{eq:lorentzian}]. This behavior is mitigated when a constant background contribution $\chi_0$ is added to the IP susceptibility, which limits the frequency range at which the polariton can be excited. Our results remain valid in the regime where the influence of $\chi_0$ is negligible, \textit{i.e.}, $f_{\parallel,\perp}\omega_0/\gamma \gg \chi_0$. 

\begin{figure}[ht]
    \centering
    \includegraphics{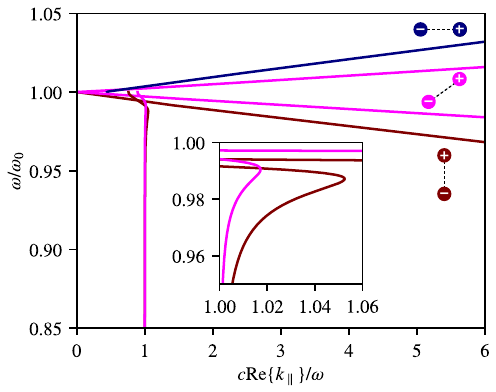}
    \caption{
    Surface polariton modes of excitons in a free-standing uniaxial sheet, with  dipoles at IP, oblique, and OOP transitions, for real frequency $\omega$ and complex wavenumber $k_\parallel$.  The free-photon-like mode is highlighted in the inset. Parameters: $\omega_0=2.1$\,eV, $\gamma=0.02$\,eV and $f_0=2$\,nm.
    }
    \label{fig:poles}
\end{figure}

To further examine these surface modes, we solve directly for the poles of the reflection coefficient, which, for a sheet suspended in air, yields ${\kappa_\perp \chi_\parallel(\omega)}={-1}$ and ${k_\parallel^2 \chi_\perp(\omega)}={\kappa_\perp}$, {with $\kappa_\perp=ik_{1,\perp}=ik_{2,\perp}$}. The corresponding modes are {obtained after isolating $\kappa_\perp$, substituting the result on} ${k_\parallel}(\omega)=\sqrt{\omega^2/c^2+{\kappa_\perp^2}}$ {and solving the resultant equation for $k_\parallel$}. We present the resulting $\omega$ vs. $\Re\{{k_\parallel}\}$ curves in Fig.~\ref{fig:poles}. We considered real $\omega$ and, consequently, complex ${k_\parallel}$. For physical meaningfulness, the results were filtered for $\Re\{{\kappa_\perp}\} > 0$, keeping the evanescent behavior away from the sheet. {Two polaritonic branches close to the exciton line are obtained, and a third} additional branch {that} approaches a free-photon mode, with low confinement since $c{\Re\{k_\parallel\}}/\omega\sim 1$ {and, thus, $\kappa_\perp\sim 0$}; hence, it does not form a properly bound surface polariton mode. We further verified the low confinement through the $\omega$ vs. $\Im\{{k_\parallel}\}$ curves, {asserting that the free-photon-approaching branch presents lower $\Im\{{k_\parallel}\}$ than the exciton-approaching branch below the exciton line across the entire spectrum}. In fact, taking into account the group velocity $v_g=d\omega/d{k_\parallel}$, we verified that$|\Im\{{k_\parallel}\}d\omega/d{k_\parallel}| < \gamma/2$ for {the free-photon-approaching branch} whereas $|\Im\{{k_\parallel}\}d\omega/d{k_\parallel}| > \gamma/2$ for {the exciton-approaching branch below the exciton line}, thereby confirming the low {confinement of the free-photon-approaching} mode.
{The discontinuity due to an out-of-plane dipole is proportional to the gradient of the out-of-plane polarization. Thus, higher wavenumber modes will possess a higher discontinuity, therefore splitting away from the dipole oscillating frequency $\omega_0$. The energy is lower for out-of-plane polaritons due to the alternating dipole configuration that it is possible to achieve, thus implying a negative group velocity. This is similar to what happens in double layer graphene, where the acoustic plasmon-polariton mode, with out-of-phase charge accumulation between neighbouring layers, possesses a lower energy than the optical (in-phase) surface mode \cite{Goncalves2016}. More details about the surface modes are explored in Section 8 of Supplement 1.}

\subsection{Curvature effects}\label{sec:reflection.curvature}

\begin{figure}[ht]
    \centering
    \includegraphics{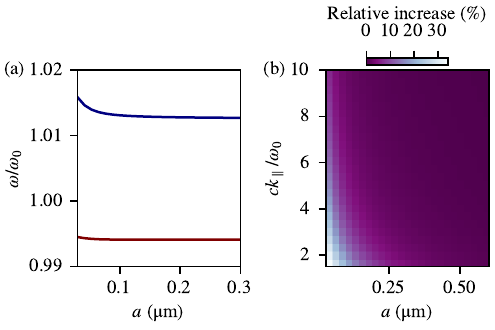}
    \caption{Surface electromagnetic modes in a cylindrical sheet of oblique dipoles. \pnl{a}~Dispersion relation as function of the cylinder radius $a$ for a fixed in-plane momentum
    $k_\parallel = 48.5\,\mathrm{\mu}\mathrm{m}^{-1}$. \pnl{b}~Relative increase of the band splitting $\omega_+-\omega_-$ as function of $k_\parallel$ and $a$. Parameters: $\omega_0=1.97$\,eV, $\gamma=0$, $\epsilon_1=1$, $\epsilon_2=2.25$, $f_0 = 2$\,nm,  and $\theta_\mathrm{dip}=20^\circ$.
    }
    \label{fig:cylinder}
\end{figure}

Oblique dipoles are poorly visible in reflectance experiments due to in-plane momentum mismatch with the surface polariton modes. As we show in a subsequent section, near-field techniques can bypass this by accessing the high-$k_\parallel$ region part of the spectra, where, as shown in the loss function spectra of Fig.~\ref{fig:loss_function}, the peaks associated with the excitation of surface modes of IP and OOP dipole components split as $k_\parallel$ increases. However, one limitation of such technique is the broad wavenumber response, i.e., not only the high-$k_\parallel$ part of the spectrum is excited, but also the low-$k_\parallel$ part, which diminishes the visibility. In Fig.~\ref{fig:cylinder}, we show the surface modes (polaritons) propagating along the axial direction in a dielectric cylinder covered with an excitonic dipole sheet. In Fig.~\ref{fig:cylinder}~\pnl{a}, we show the dependence on the cylinder radius for a fixed $k_\parallel=48.5\,\mu\mathrm{m}^{-1}$. As the radius decreases, both the IP and OOP associated branches increase in energy, with a net increase in the splitting between the bands. In panel~\pnl{b}, the net increase in the split of the bands is plotted as a function of both the wavenumber and the cylinder radius, showing that the relative increase is higher for lower momentum and cylinder radius. Thus, besides showing an application of this formalism for a curved geometry, we also obtained that geometry can be used to increase the effects of oblique dipoles in near-field measurements. The higher splitting between the two bands can facilitate the measurement of oblique dipoles.

Further details on the application of the BCs~\eqref{eq:BCs} to the curved geometry here discussed are presented in Section 9 of Supplement 1.

\section{Near-field characterization of uniaxial dipole sheets}\label{Sec4}

Near-field spectroscopy allows access to physical properties at sub-wavelength resolution. Several experimental techniques have explored this regime of electrodynamics, with s-SNOM being one of the most successful ones~\cite{chen2019modern}. It consists of shedding a focused laser source on the material under investigation, with a metallic tip in its proximity. The field back-scattered due to laser incidence on the tip-sheet system will then be measured by a detector in the far-field region (relative to the tip and sheet). By applying a mechanical oscillation to the tip at a frequency $\Omega$, the back-scattered field is modulated in time with components at integer multiples of $\Omega$. These components are probed by a detector in the far-field region, and signatures of tip-sheet interaction in the near-field region can be identified through peaks within the demodulated field measured in the far-field region.

For a tip-sheet distance ${h}(t)$ oscillating at frequency $\Omega$, the near-field is $E_\mathrm{NF}[{h}(t)]=\sum_{n=-\infty}^{\infty} \sigma_n e^{-in\Omega t}$, where
\begin{equation}
    \sigma_n = \frac{1}{T}\int_0^T dt~e^{in\Omega t}E_\mathrm{NF}[{h}(t)],\label{eq:Fourier_coefficients_of_the_near-field}
\end{equation}
and $T=2\pi/\Omega$ is the period of oscillation. Several techniques have been developed to extract the Fourier coefficients $\sigma_n$ from the signals probed by the detector in an s-SNOM setup~\cite{Dai2018}. Signatures of near-field interaction are then identified by peaks in the modulus and phase of $\sigma_n$.

Describing the near-field interactions of tip and sample in an s-SNOM setup is a challenging task, which has been tackled in works that mix analytical, numerical, and machine learning techniques~\citep{Li1997,knoll2000enhanced,Hillenbrand2001,Walford2001,cvitkovic2007analytical,aizpurua2008substrate,hauer2012quasi-analytical,Gao2018,chen2021,Nayak2021,robledo2025theoretical,Garrity2025,voronin2025quantitative}.

The most successful attempts at an analytical model---or, at least, semi-analytical---treat the tip as a nanosphere or nanospheroid. The former has a charge distribution that effectively behaves as a point-dipole located at the sphere's center; the latter, as a finite-dipole formed by a pair of point-charges at different positions of the spheroid's major axis. For a comparison between the point-dipole model (PDM) and the charge averaged finite-dipole model (FDM) \cite{Mester2020}, the reader is referred to Section 6 of Supplement 1. In what follows, we describe the s-SNOM tip and its interaction with the dipole sheet using a customized version of the PDM, which retains retardation effects by allowing the Green function of tip-sheet interaction to be solved numerically. Our main results in the following sections are a consequence of the split in momentum from the two polaritonic branches of oblique dipoles as shown in Fig. \ref{fig:loss_function} and the capacity of near-field techniques such as SNOM to excite those high-wavenumber modes. Thus, the qualitative results are robust against the specificities of the model used to describe the tip-substrate interaction.

\subsection{Point-Dipole Model}\label{Sec4.1}

We consider a metallic tip brought close to the sheet as in Fig.~\ref{fig:optical_configurations}\pnl{b}, aligned along the surface normal. The elongated tip geometry results in a dominant polarization towards the OOP direction. The dipole induced in the tip is modeled in the PDM as the response of a sphere of polarizability $\alpha(\omega)$. The OOP component of the total field acting on the tip may be expanded in terms of an external contribution $E_{0,{\perp}}$ and a contribution coming from the dipole's radiation reflected back into itself:
\begin{subequations}
\begin{eqnarray}
    E_{{\perp}} = E_{0,{\perp}} + G p, \label{eq:total field on the tip}
    \\
    p = \alpha E_{{\perp}}. \label{eq:p=alphaE 0}
\end{eqnarray}
\end{subequations}
Additionally, for a linear homogeneous medium in the absence of additional sources, the external field may be written generically in terms of plane waves propagating in the $+\hat{\mathbf n}$ and $-\hat{\mathbf n}$ directions. For concreteness, we express this external field as a monochromatic laser, with wavevector $\mathbf k_{m}^{\mathrm{inc}} = {\mathbf k_{\parallel}^{\mathrm{inc}}}-{k_{m,\perp}^{\mathrm{inc}}\hat{\mathbf n}}$. An additional contribution is generated from direct reflection into the sheet before scattering in the tip. Then,
\begin{equation}
    E_{0,{\perp}} = E_{\mathrm{inc},{\perp}} + r_p E_{\mathrm{inc},{\perp}} e^{2i{k_{\perp,\mathrm{inc}}}{h}}, \label{eq:expanded external field}
\end{equation}
Substituting Eqs~(\ref{eq:total field on the tip},\ref{eq:expanded external field}) in Eq.~\eqref{eq:p=alphaE 0} results in
\begin{equation}
    p = \alpha E_{\mathrm{inc},{\perp}} \left(1 + r_p e^{2i{k_{\perp,\mathrm{inc}}}{h}}\right) + \alpha G p. \label{eq:p=alphaE}
\end{equation}

Before proceeding, we emphasize the assumptions being made when writing the tip's dipole moment as in Eq.~\eqref{eq:p=alphaE}: The dipole moment {of the tip} is restricted to {surface normal (OOP)} direction as a phenomenological correction of the model to accommodate for the elongated tip geometry. The last term, $\alpha G p$, includes all tip scattering events, whereas the remaining terms are due to field contributions that exist with or without the presence of the tip. In the following, we proceed to rewrite an expression for the {tip's} dipole moment solely in terms of the incident field $E_{\mathrm{inc},{\perp}}$ and an effective polarizability $\alpha_\mathrm{eff}$, which renormalizes the tip to account for the presence of the sheet.

The polarizability may be written in the Clausius--Mossotti form
\begin{equation}\label{eq:alpha Clausius-Mossotti}
    \alpha(\omega) = 4\pi\epsilon_0 R_\mathrm{tip}^3 \frac{\epsilon_\mathrm{tip}(\omega)-1}{\epsilon_\mathrm{tip}(\omega) + 2},
\end{equation}
for a radius of curvature $R_\mathrm{tip}$ and a tip permittivity $\epsilon_\mathrm{tip}(\omega)$~\cite{Amorim2017}. The factor $(1+{r_p}e^{2i{k_\mathrm{\perp,inc}}{h}})$ includes the reflection of the external field on the sheet. The Green's function $G =\omega^2\mu_0 G_{{\perp\perp}}(\mathbf r',\mathbf r',\omega)$ results from self-reflection of the field radiated by the dipole back into itself. Equation~\eqref{eq:p=alphaE} can be rewritten as $p=\alpha_\mathrm{eff} \left(1+r_pe^{2ik_\mathrm{\perp,inc}{h}}\right)E_\mathrm{inc,{\perp}}$, where
\begin{equation}
    \alpha_\mathrm{eff} = \frac{\alpha}{1-\alpha G} = \frac{1}{\alpha^{-1} - G}\label{eq:polarizability}
\end{equation}
is the renormalized polarizability. An expression for $G$ may be derived from the solution of Maxwell's equations for a point-dipole with BCs at the sheet interface (see {Section 3 of the Supplement 1}):
\begin{equation}
    G = \frac{i}{4\pi\epsilon_0} \int_0^\infty d{k_\parallel} \frac{{k_\parallel}^3}{{k_{1,\perp}}} e^{2i{k_{1,\perp}} {h}} {r_p}({k_\parallel},\omega).\label{eq:G integral}
\end{equation}

The back-scattered field is given by the sum of the point-dipole radiation with its reflection on the sheet. Up to leading order, one finds (see {Section 3 of the Supplement 1})
\begin{equation}
    E_\mathrm{sca}(\mathbf R\rightarrow\infty,\omega) = \frac{ip_\mathrm{tot}ke^{ikR}}{2\epsilon_0 R},
\end{equation}
where $k=\omega/c$ and $p_\mathrm{tot}({h})=p\left(1 + r_p({\mathbf k_{\parallel,\mathrm{inc}}},\omega)e^{2ik_\mathrm{\perp,inc}{h}}\right)$, to make sure we include both the reflection of the external field and the reflection of the point-dipole's field on the sheet, as pointed out by Ref.~\cite{cvitkovic2007analytical}. We can write
\begin{equation}
    p_\mathrm{tot}({h},\omega) = \alpha_\mathrm{eff}({h},\omega)E_\mathrm{inc,{\perp}}\left(1+r_p({\mathbf k_{\parallel,\mathrm{inc}}},\omega)\right)^2 + \mathcal{O}(k_{{\perp},\mathrm{inc}}{h})\label{eq:total dipole moment representing the scattered field}
\end{equation}
and the Fourier coefficients~\eqref{eq:Fourier_coefficients_of_the_near-field} are then given by
\begin{equation}
    \sigma_n = \frac{ike^{ikR}E_\mathrm{inc,{\perp}} \left(1+r_p(\mathbf k_{\parallel,\mathrm{inc}},\omega)\right)^2}{2\epsilon_0 R T} \int_0^T dt~e^{in\Omega t}\alpha_\mathrm{eff}[{h}(t)].\label{eq:snom}
\end{equation}

The signal measured in an s-SNOM detector, by means of proper demodulation techniques~\cite{Dai2018}, is proportional to the harmonics $\sigma_n$ of the back-scattered field, given by Eq.~\eqref{eq:snom} in our PDM. To filter out global phases and unimportant scaling factors, the signal can be normalized as
\begin{equation}
    \bar{\sigma}_n =\frac{\sigma_n}{\sigma_{n}^{\mathrm{subs}}} = \dfrac{\left(1+r_p(\mathbf k_{\parallel,\mathrm{inc}},\omega)\right)^2}{\left(1+r_p^\mathrm{subs}(\mathbf k_{\parallel,\mathrm{inc}},\omega)\right)^2}\frac{\int_0^T dt~e^{in\Omega t} \alpha_\mathrm{eff}[{h}(t)]}{\int_0^T dt~e^{in\Omega t} \alpha_\mathrm{eff}^\mathrm{subs}[{h}(t)]} ,\label{eq:normalized s-SNOM harmonics}
\end{equation}
where $r_p^\mathrm{subs}$ is the reflection coefficient of the bare substrate, \textit{i.e.}, Eq.~\eqref{eq:Fresnel reflection coefficient for p-polarized light} particularized to $\chi_\parallel=\chi_\perp=0$. Accordingly, the substrate effective polarizability $\alpha_{\mathrm{eff}}^{\mathrm{subs}}$ is obtained by setting $r_p = r_p^{\mathrm{subs}}$ on Eq.~\eqref{eq:polarizability}.


\subsection{Oblique dipole moments} \label{Sec4.3}

As discussed in Sec.~\ref{sec:reflection},

the excitonic uniaxial sheet has poor visibility of the oblique dipole moment feature in far-field reflectivity. In the following, we investigate how the enhanced OOP field-dipole coupling caused by a tip in the near-field region affects the optical response. For simplicity, we consider a dipole sheet with a single excitonic transition at $\omega_0$. The tip characterization follows conventions of the PDM for reproduction of s-SNOM experiments in the VIS range: the tip radius is $R_\mathrm{tip}=25~\mathrm{nm}$, and the permittivity $\epsilon_\mathrm{tip}(\omega)$ entering the tip's polarizability~\eqref{eq:alpha Clausius-Mossotti} is the dielectric function of platinum, taken from experimental data in a range of $0.1$--$5~\mathrm{eV}$~\cite{Rakic98_Optical-properties}. The tip-sheet distance entering the Green function~\eqref{eq:G integral} oscillates following ${h}(t) = h_0 + A[1-\cos(\Omega t)]$, with $A\approx 50$~nm being the widely accepted amplitude value~\cite{aizpurua2008substrate,robledo2025theoretical,Zhang2022}, whereas $h_0$ can be varied freely within $h_0 > R_\mathrm{tip}$. Smaller distances $h_0$ in general mean stronger tip-sample coupling and higher-contrast peaks in the near-field spectra~\cite{cvitkovic2007analytical}, however as ${h}(t)\to R_\mathrm{tip}$ the Green function~\eqref{eq:G integral} diverges. We therefore set a minimum distance of $1~\mathrm{nm}$ between the sheet and the tip apex ($h_0=26~\mathrm{nm}$), following the community convention~\cite{robledo2025theoretical,aizpurua2008substrate,Vincent2024snompy}.

\begin{figure*}[ht]
    \centering
    \includegraphics{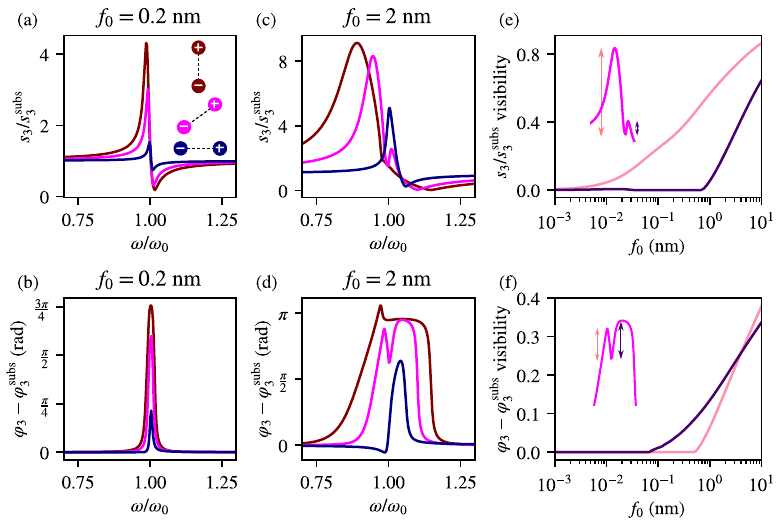}
    \caption{Near-field spectroscopy of excitons in a uniaxial sheet with a single transition at $\omega_0$. \pnl{a,b}~Modulus ($s_3$) and phase ($\varphi_3$) of the $n=3$ normalized field harmonics~[Eq.~\eqref{eq:normalized s-SNOM harmonics}] for an oscillator strength $f_0=0.2$\,nm and dipole angles $\theta_\mathrm{dip}=\frac{\pi}{2}$ (OOP), $\theta_\mathrm{dip}=45^\circ$ (oblique), and $\theta_\mathrm{dip}=0$ (IP). \pnl{c,d}~Same as panels \pnl{a,b}, but with increased oscillator strength ($f_0=2$\,nm). \pnl{e,f}~Visibility, defined as $(P-D)/(P+D)$, of the peaks $P$ at $\omega<\omega_0$ and $\omega>\omega_0$ relative to the dip $D$ at $\omega_0$. Parameters: $\omega_0=2.1$\,eV, $\gamma=0.02$\,eV, $\epsilon_1=1$, $\epsilon_2=2.25$, $R_\mathrm{tip}=25$\,nm, $h_0=26$\,nm, and $A=50$\,nm.}
    \label{fig:near_field_oblique}
\end{figure*}

The near-field spectrum of the dipole sheet, with Green's function integrated numerically (see Supplement 1, Sec.~5), presents a peak close to $\omega_0$ [Fig.~\ref{fig:near_field_oblique}\pnl{a}], indicating the excitation of the surface polariton modes discussed in~Sec.~\ref{sec:reflection} through access of momenta beyond the light line. To allow for direct comparison with the far-field characterization in Sec.~\ref{sec:reflection}, the sheet was placed on top of glass ($\epsilon_2=2.25$). Additionally, the incidence angle does not affect the near-field spectrum significantly (see {Supplement 1, Section 3}), therefore we fix normal incidence ($\theta_\mathrm{inc}=0^\circ$) for simplicity. By sweeping the dipole orientation from OOP down to oblique and, ultimately, IP orientation, the peak decreases monotonically both in modulus and in phase, a consequence of the selectively enhanced coupling to the OOP component.

As the oscillator strength is increased, the peak on modulus and phase spectra splits into two, indicating surface polariton excitation at frequencies further away from $\omega_0$.
These splittings of the lineshape into two peaks are signatures of oblique dipole moment, with significantly improved visibility [Fig.~\ref{fig:near_field_oblique}\pnl{e,f}], verified by the large visibility values obtained at significantly lower oscillator strengths when compared to the far-field technique [Fig.~\ref{fig:reflection}\pnl{e} for $\omega_{0\perp}\approx\omega_{0\parallel}=\omega_0$].

As a follow up, we would like to bring attention to the shape of the $s_3$ curves in Figs.~\ref{fig:near_field_oblique}\pnl{a} and~\pnl{c}.
The asymmetric lineshapes in near-field spectra, as punctuated in related works~\cite{Bortchagovsky2002,aizpurua2008substrate}, indicate an interference between a sharp and a broad resonance in analogy with that of a discrete state with a continuum of states in quantum mechanics. The tip plus substrate supports a broad range of modes, extending in momentum up to $q\sim 1/R_\mathrm{tip}$, which is the continuum of states that couples with the OOP dipole moment, due to close-to-vertical near-field in the tip apex. For more details on the Fano-like shapes of the near-field spectra, the reader is referred to Section 4 of the Supplement 1.

\subsection{Dark exciton in monolayer WSe$_2$}\label{Sec4.2}

\begin{figure}[ht]
    \centering
    \includegraphics{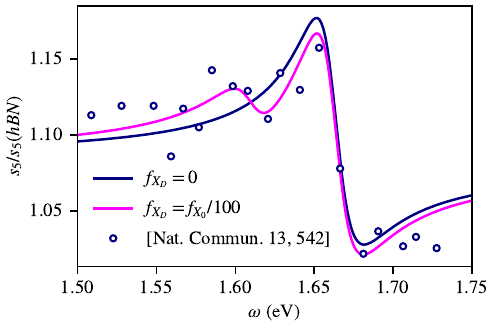}
    \caption{Secondary peak emergence in the near-field spectra of monolayer $\mathrm{WSe}_2$ due to a detuned dark exciton. Continuous curves represent PDM predictions, and dots, experimental data from Ref.~\cite{Zhang2022}. Parameters: $\omega_0=1.66$\,eV, $\omega_D=1.61$\,eV, $\gamma_{r,0}=1.4$\,meV, $\frac{\gamma_{nr}}{2}+\gamma_d=15$\,meV, $\epsilon_\infty=9.7$, $d=0.7$\,nm, $R_\mathrm{tip}=25$\,nm, $h_0=26$\,nm, and $A=50$\,nm.}
    \label{fig:experiment}
\end{figure}

Due to the scarcity of near-field experiments reporting oblique dipole moments in the literature (to the best of our knowledge), to benchmark our customized PDM against available near-field experimental data we discuss monolayer WSe$_2$ in this Section, a system that hosts a bright IP mode and a dark OOP mode~\cite{Zhang2022}.
Figure~\ref{fig:experiment} shows the near-field spectra of monolayer $\mathrm{WSe}_2$, incorporated in our model through the susceptibility
\begin{subequations}\label{eq:chi parallel TMD}
\begin{align}
    \chi_{1,\parallel}(\omega) &= \chi_{2,\parallel}(\omega) = \frac{d}{2}\left[\epsilon(\omega)-1\right], \label{eq:chi parallel TMD internal}
    \\
    \epsilon(\omega) &= \epsilon_\infty - \frac{c}{d\omega_0} \frac{\gamma_{r,0}}{\omega-\omega_0+i\left(\frac{\gamma_{nr}}{2}+\gamma_d\right)},
\end{align}
\end{subequations}
for a bright exciton transition at $\omega_0$ with radiative decay rate $\gamma_{r,0}$, non-radiative decay rate $\gamma_{nr}$, and dephasing decay rate $\gamma_d$, on a monolayer of thickness $d$ and background permittivity $\epsilon_\infty$. The resulting near-field spectrum for $\chi_{1,\perp}=\chi_{2,\perp}=0$ is presented in Fig.~\ref{fig:experiment} for the bright exciton parameters extracted from Ref.~\cite{Zhang2022}, showing nearly identical qualitative agreement with the experimental data. The background susceptibility, represented by $d(\epsilon_\infty-1)$ in Eq.~\eqref{eq:chi parallel TMD internal}, was verified to shift the resulting lineshape vertically in the spectral range shown in Fig.~\ref{fig:experiment}. In other words, a constant contribution to the sheet's IP susceptibility affects mostly the quantitative aspect on the lineshape. Nonetheless, for the intent of matching the PDM to experimental data, keeping the background contribution to the susceptibility was essential.

To include a dark exciton transition at $\omega_D$, which has an OOP dipole moment, we put
\begin{equation}\label{eq:chi perp TMD}
    \chi_{1,\perp}(\omega) = \chi_{2,\perp}(\omega) = \dfrac{f_{X_D} \omega_D^2}{\omega_D^2-\omega^2-i\gamma_D\omega},
\end{equation}
with oscillator strength $f_{X_D}$ and damping $\gamma_D$ to be determined. To tie this formulation to how the optical response of TMDs is usually presented in the literature, we can use $(\omega_D^2-\omega^2-i\gamma_D\omega)^{-1} \approx (2\omega_D)^{-1}(\omega_D-\omega-i\gamma_D/2)^{-1}$ (valid at optical frequencies) to write
\begin{equation}
    \dfrac{f_{X_D}\omega_D^2}{\omega_D^2-\omega^2-i\gamma_D\omega} \approx -\dfrac{\omega_D}{2} \dfrac{f_{X_D}}{\omega-\omega_D+i\frac{\gamma_D}{2}},
\end{equation}
from which we find the relation $\gamma_r = \omega^2 f c^{-1}$ between the radiative decay rate $\gamma_r$ and the oscillator strength $f$ of an excitonic optical transition at $\omega$, by direct comparison with Eqs.~\eqref{eq:chi parallel TMD}.

Figure~\ref{fig:experiment} showcases the near-field spectra obtained setting $f_{X_D}=f_{X_0}/100$ (with $f_{X_0}$ defined as $f_{X_0}=2c\gamma_{r,0} \omega_0^{-2}$) for a dark exciton redshifted by $50~\mathrm{meV}$ relative to the bright counterpart, with equal damping ($\gamma_D=\frac{\gamma_{nr}}{2}+\gamma_d$). A secondary spectral feature arises, indicating a greater qualitative agreement with the experimental data from Ref.~\cite{Zhang2022}. The chosen bright-dark exciton detuning is in agreement with several literature reports~\cite{Zhang2017MagneticBrightening,Zhou2017,Park2018}. The radiative decay rate of the dark exciton is still subject to debate, however some studies claim that $\gamma_{r,D}$ is a couple orders of magnitude smaller than $\gamma_{r,0}$~\cite{Zhang2017MagneticBrightening,Slobodeniuk2016SpinFlipProcesses}. Moreover, the exact form of the OOP component of the constitutive relation between $\mathbf{P}^\mathrm{2D}$ and $\mathbf{E}$ is also in question~\cite{Hansen2026}, \textit{i.e.}, additional multiplicative factors relating $f_{X_D}$ and $f_{X_0}$ could be at play as was emphasized in the ending of Sec.~\ref{sec:boundary_conditions}. Nonetheless, Fig.~\ref{fig:experiment} shows a greater qualitative agreement for $f_{X_D}=f_{X_0}/100$ rather than $f_{X_D}=0$, treated here as the net oscillator strength of the dark exciton subject to eventual screening effects. Additionally, the root mean square residual of the fittings showed an improvement larger than $20\%$ for the oblique case relative to the IP case, quantitatively corroborating the agreement with the experimental data.

We verified that changing the dark exciton oscillator strength to $f_{X_D}=f_{X_0}/1000$ washes out the second peak almost completely. This large sensitivity may be understood in terms of two phenomena: first, the elongated shape of the tip along the OOP direction enhances coupling with optical transitions possessing OOP dipole moments~\cite{Zhou2017}, an effect which is emulated in the PDM by restricting the total dipole moment [Eq.~\eqref{eq:total dipole moment representing the scattered field}] to the OOP direction; second, the optical transitions with IP dipole moment are screened due to the encapsulating layer of hBN, an effect which can be verified through the reflection coefficient $r_p$ [Eq.~\eqref{eq:Fresnel reflection coefficient for p-polarized light}]. The in-plane coefficient $\beta_{2,p,\parallel}$ shows a factor of $1/\epsilon_2$, which is not present in the OOP coefficient $\beta_{2,p,\perp}$.

The choices made in the definition of $f_{X_0}$ and $f_{X_D}$ correspond, in our former $\{f_\parallel,f_\perp\}$ formulation [Eq.~\eqref{eq:symmetric_oscillator_strengths}], to $
    \frac{f_{X_D}}{f_{X_0}} = \frac{f_\perp}{2f_\parallel} = \tan^2(\theta_\mathrm{dip}).
$
For $f_{X_D}=f_{X_0}/100$ as in Fig.~\ref{fig:experiment}, one gets $\theta_\mathrm{dip}\approx 5.7^\circ$.

\section{Conclusion}\label{Sec5}

To account for the out-of-plane dipole orientation, {we have developed a general formalism for polarization confined to a surface of deeply subwavelength thickness, which applies to} Felderhof's formalism of linear polarization sheets {and to Feibelman's SRF of metal-dielectric interfaces}. By doing so, we developed a flexible model that provides a unified theoretical language. It applies to dipolar sheets{, which arise from excitons in 2D materials and layers of strongly polarizable molecules, plasmons in metal-dielectric interfaces, and possibly many other systems where the dipolar contribution to charge density comes at leading order.} Our formalism provides {a zero-thickness surface treatment of dipole polarization confined to regions of atomic-scale thickness, aiding the transparency of the underlying physics as compared to the thin-film model, and advancing into the understanding of quasi-2D structures coming from polarization stemming out of the plane.}

We demonstrated the characterization of average dipole orientation in a layer of excitonic transition dipoles confined to two-dimensions. By examining a planar sheet on a substrate of weak reflectivity, we were able to show distinct reflectance peaks clearly associated with exciton-polaritons originating from in- and out-of-plane dipole contributions, {which form a double peak spectral feature at oblique dipole orientation}. These peaks achieve higher visibility when employing near-field techniques--- here exemplified through an s-SNOM setup---, which generates Fano-like resonances that capture the effect of surface modes in the sheet. {The elongated form of the metallic tips usually employed in near-field spectroscopy is particularly valuable for selectively enhancing the coupling with OOP resonant features, as demonstrated with monolayer $\mathrm{WSe}_2$. Further efforts on the experimental characterization of OOP near-field response are essential for disambiguating the current literature and converging on a unified theoretical description.}

\begin{backmatter}
\bmsection{Funding}
São Paulo Research Foundation (2023/14231-6, 2024/05040-5);
Independent Research Fund Denmark (2032-00045B).
Portuguese Foundation for Science and Technology (UID/04650/2025, PTDC/FIS-MAC/2045/2021).
Danish National Research Foundation (Project No.~DNRF165).

\bmsection{Acknowledgments}
V.~G.~M.~D. acknowledges a PhD. Scholarship from the São Paulo Research Foundation (FAPESP), Processes No.~2023/14231-6 and~2024/05040-5.
N.~M.~R.~P. and N.~A.~M. acknowledge the Independent Research Fund Denmark (grant no. 2032-00045B).
N.~M.~R.~P. also acknowledges support from the Portuguese Foundation for Science and Technology (FCT) by the project PTDC/FIS-MAC/2045/2021.
The Center for Polariton-driven Light--Matter Interactions (POLIMA) is funded by the Danish National Research Foundation (Project No.~DNRF165). 
This work was supported by the Portuguese Foundation for Science and Technology (FCT) in the framework of the Strategic Funding UID/04650/2025.

\bmsection{Disclosures}
The authors declare no conflicts of interest.

\bmsection{Data Availability}
All data supporting the findings of this study are available within the article.

\bmsection{Supplement 1}
See Supplement 1 for supporting content.

\end{backmatter}

\bibliography{references.bib}

\ifarXiv
    \foreach \x in {1,...,\numbersupplementpages}
    {
        \clearpage
        \includepdf[pages=\x]{\supplementfilename}
    }
\fi

\end{document}